\documentstyle[11pt,moriond,epsfig]{article}

\def\Journal#1#2#3#4{{#1} {\bf #2}, #3 (#4)}

\def\NIMA{{\em Nucl. Instrum. Methods} A}

\def\PLB{{\em Phys. Lett.}  B}
\def\PRL{\em Phys. Rev. Lett.}
\def\PRD{{\em Phys. Rev.} D}

\def\EPJC{{\em Eur. Phys. J.} C}
\def\CPC{\em Comp. Phys. Comm.}

\def\ra{\rightarrow}

\def\be{\begin{equation}}
\def\ee{\end{equation}}
\def\bea{\begin{eqnarray}}
\def\eea{\end{eqnarray}}

\def\kT{\langle k_T \rangle}
\def\kTintr{\langle k_T^{\rm intr} \rangle}
\def\GeV{\rm \,GeV}
\def\rad{\rm \,rad}

\begin{document}
\vspace*{4cm}
\title{PROMPT PHOTONS AND DVCS AT HERA}

\author{J.C.\,HART}

\address{Rutherford Appleton Laboratory, Chilton, DIDCOT,
Oxfordshire  OX11 0QX, England\\
(on behalf of the ZEUS and H1 collaborations)}

\maketitle\abstracts{
A recent analysis of prompt-photon events in ZEUS has led to a new
determination of the effective transverse parton momentum in the proton.
The observation of deeply virtual Compton scattering by ZEUS and H1,
including a first measurement of the DVCS cross section at HERA by
H1, is also reported.}

\section{Introduction}

The production of real photons in high energy $ep$ collisions (prompt photons)
is of interest because it allows the parton level interactions to be 
investigated while minimising the complication of parton hadronisation.
Two recent studies of prompt-photon production at HERA are reported 
in this paper.

\begin{figure}[h]
   \centering
   \epsfig{file=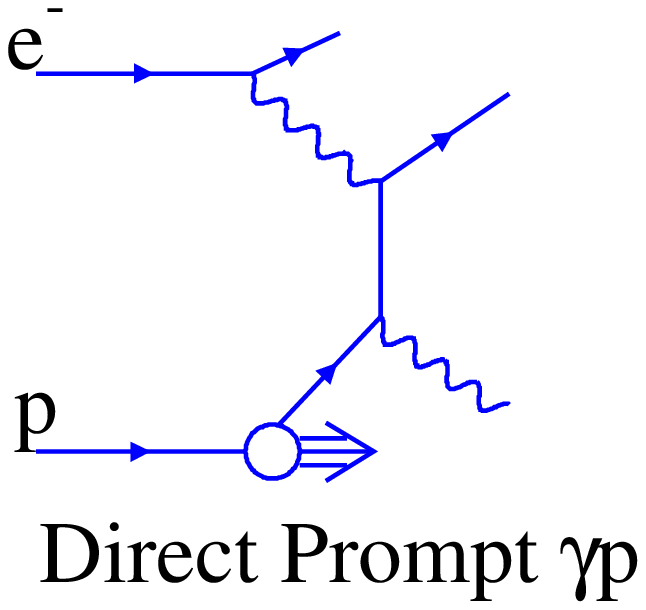,width=3.0cm}%
   \hspace{5mm}%
   \epsfig{file=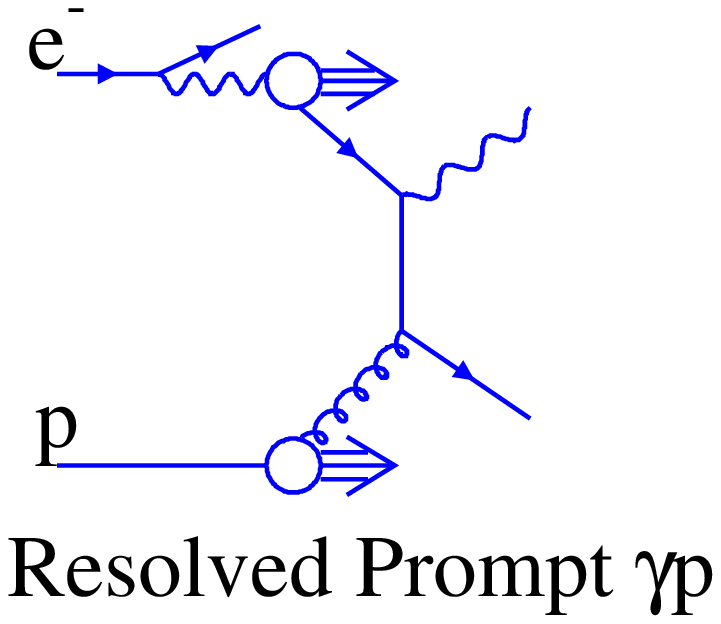,width=3.0cm}%
   \hspace{10mm}%}
   \epsfig{file=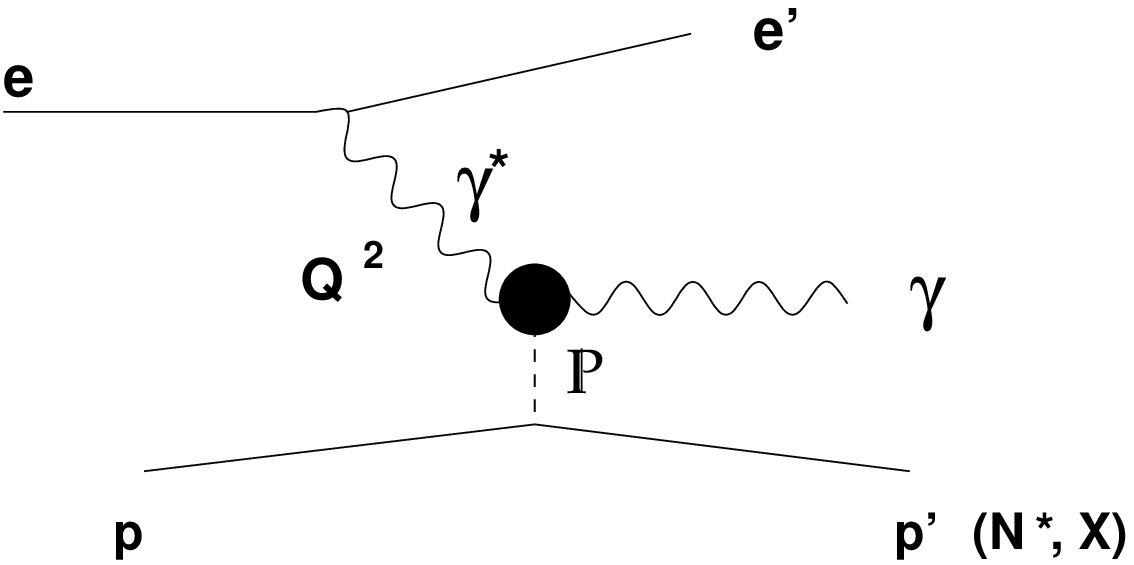,width=5.0cm}
   \caption{Typical diagrams for prompt photon production in direct
            and resolved photoproduction are shown on the left.
            The right-hand diagram is for DVCS.}
   \label{fig:intro_diagrams}
\end{figure}

The first reaction to be discussed is prompt-photon photoproduction 
where a real photon is produced directly from the hard interaction of 
a quasi-real photon. 
Both direct processes (where the entire photon participates in the 
interaction) and resolved processes (where the photon first fluctuates
into a hadronic system) are possible as shown by the examples in 
Fig.\,\ref{fig:intro_diagrams}.
ZEUS has used the direct process to determine $\kT$, the 
effective transverse parton momentum in the proton.

The second process is deeply virtual Compton scattering (DVCS) in
which a real photon is produced through the diffractive scattering
of a virtual photon off the proton as shown in Fig.\,\ref{fig:intro_diagrams} 
(right-hand diagram).
DVCS makes it possible to study exclusive diffraction without the 
theoretical uncertainty introduced by the vector meson wave function.
It is also sensitive to the skewed parton distributions which 
describe two-parton correlations in the proton.
DVCS has been seen at HERA by both ZEUS and H1, and H1 has measured
differential cross-sections.

\section{Prompt photons in photoproduction and determination of $\kT$}

The ZEUS analysis of prompt photons is based on data taken at the 
HERA e-p collider in 1996-97 with positron and proton beam energies
of 27.5 and 820$\GeV$, respectively. 
The ZEUS detector, including the uranium-scintillator calorimeter
and central tracking detector (CTD), which are relevant for this study,
have been described elsewhere.\cite{zeus1,cal,ctd}

The signature for prompt photon events was an isolated cluster of cells 
in the electromagnetic section of the ZEUS barrel calorimeter.
At least one visible jet was required to determine $\kT$.
Events with the scattered positron seen in the calorimeter were rejected.
This selected photoproduction events with $Q^2 \leq 1 \GeV^2$ where $Q^2$ 
is the virtuality of the incoming photon.
The other event selection cuts were
$$\begin{array}{ccc}
E_T^{\gamma}>5\GeV  &  \hspace{1cm}  &  -0.7<\eta^{\gamma}<0.9   \\
E_T^{jet}>5\GeV     &  \hspace{1cm}  &  -1.5<\eta^{jet}<1.8    
\end{array}$$
where $\eta$ is the pseudorapidity, $\eta=-\ln\tan(\theta/2)$, and
the polar angle $\theta$ is measured with respect to the proton beam.
The photon isolation requirement was the absence of any other 
calorimeter cluster in a unit $(\eta,\phi)$ cone around the photon
direction, where $\phi$ is the azimuthal angle.

Photons were distinguished from neutral mesons ($\pi^0$ and $\eta^0$)
by virtue of their smaller cluster size in the calorimeter, as in 
previous ZEUS analyses.\cite{zeus1,zeus2}
A cut of 3.25\,cm was first applied to the cluster width in the 
$z$-direction (parallel to the proton beam) to enhance the photon purity.
The fraction, $f_{max}$, of the total energy in a single calorimeter cell
peaks near 1.0 for photons, but is flatter for $\pi^0$ and $\eta^0$ 
and is insensitive to the $\pi^0/\eta^0$ ratio.
It was therefore used to make a statistical subtraction of the remaining
$\pi^0$ and $\eta^0$ background.  

The incoming photon energy for each event was estimated from 
$y_{JB}=\sum{(E-p_z)/2E_e}$, where the sum is over over all energy flow 
objects\,\cite{zufos} with energy $E$ and longitudinal momentum $p_z$ 
and $E_e$ is the incident positron beam energy.
Limits of $0.2<y_{JB}<0.7$ were applied to remove beam-gas and deep 
inelastic scattering events.
Direct photoproduction events were selected by applying a cut
$x_{\gamma}^{\rm obs}>0.9$, where
$$x_{\gamma}^{\rm obs}=\frac{1}{2E_e y_{JB}} \sum_{\gamma+{\rm jet}}(E-p_z)$$
is the estimated fraction of the incoming photon energy in the QCD subprocess.
Events passing this cut were nearly all from direct photoproduction.

\begin{figure}[h]
   \centering
   \epsfig{file=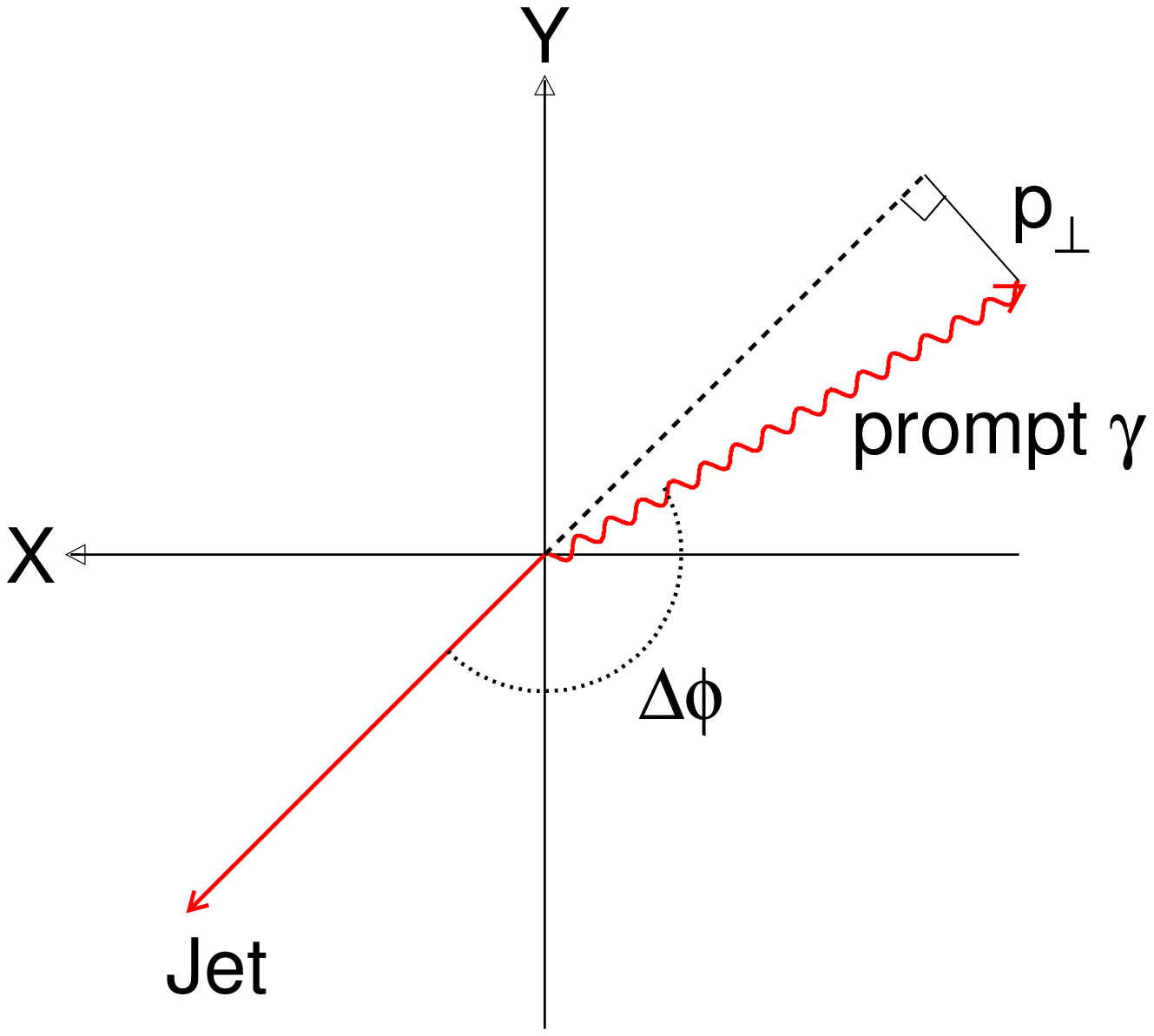,width=5cm,height=4.5cm,
           bbllx=106,bblly=76,bburx=483,bbury=415}%
   \hspace{1cm}%
   \epsfig{file=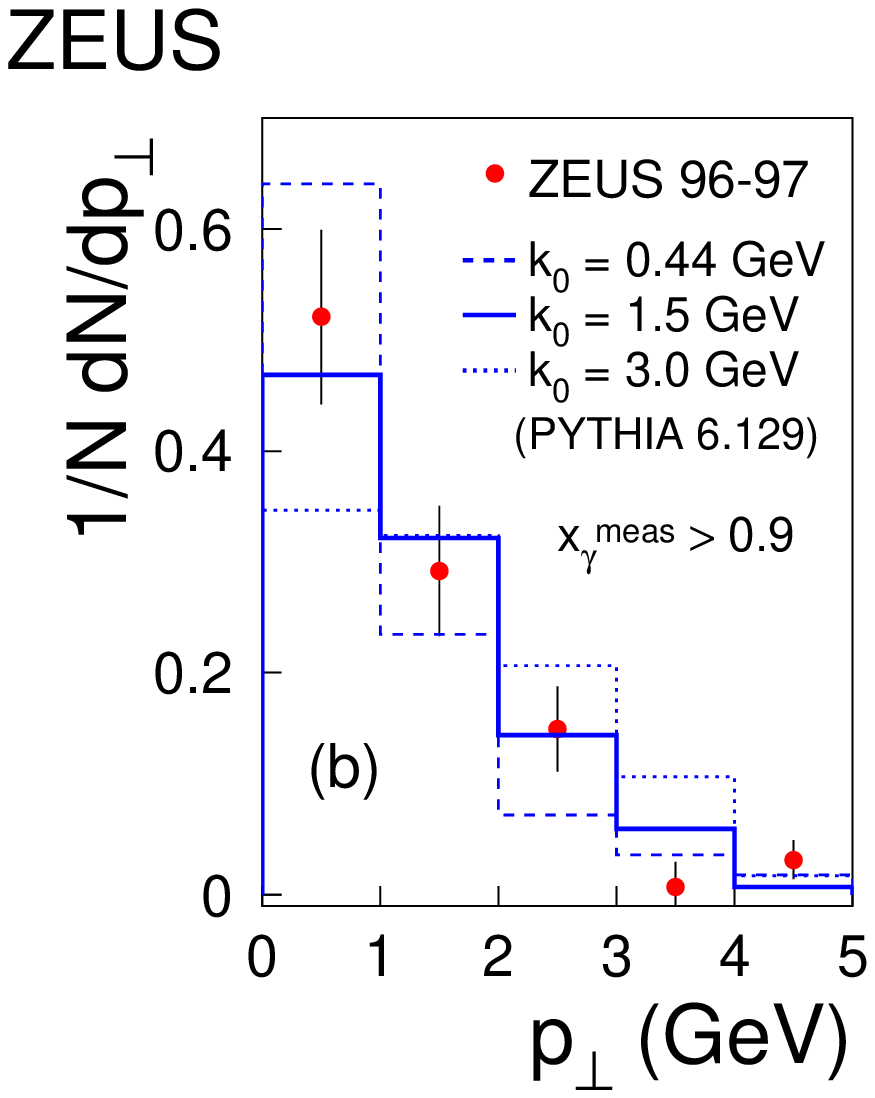,width=6.5cm,height=5.5cm,
           bbllx=267,bblly=396,bburx=515,bbury=707}%
   \caption{The left-hand diagram defines $p_\perp$.
            The X- and Y-axes are orthogonal to the proton beam.
            The distribution of $p_\perp$ is shown on the right
            and compared to the predictions of PYTHIA for
            different values of $k_0$.}
   \label{fig:p_perp}
\end{figure}

To determine $\kT$, the distribution of $p_\perp$, the momentum component
of the photon perpendicular to the jet direction (see Fig.\,\ref{fig:p_perp}), 
was compared to the predictions of PYTHIA\,\cite{PYTHIA} for different 
values of the parameter $k_0$, where $k_0=\sqrt{4/{\pi}}\kTintr$ and 
$\kTintr$ is the mean absolute value of the intrinsic parton momentum 
in the proton in the PYTHIA model.
The results may be seen in Fig.\,\ref{fig:p_perp} which shows that the data
are compatible with a value of $k_0$ between about $0.4\GeV$ and $3.0\GeV$.
A fit to the $p_\perp$ distribution was performed, with additional $k_0$ 
points, to find the optimal value of $k_0$ and hence $\kTintr$.
The corresponding value of $\kT$ was calculated by adding the parton
shower contribution obtained from PYTHIA ($\sim1.4\GeV$) to give
$$\kT=1.69\pm0.18^{+0.18}_{-0.20}\GeV.$$

\begin{figure}[h]
   \centering
   \vspace{-2cm}
   \epsfig{file=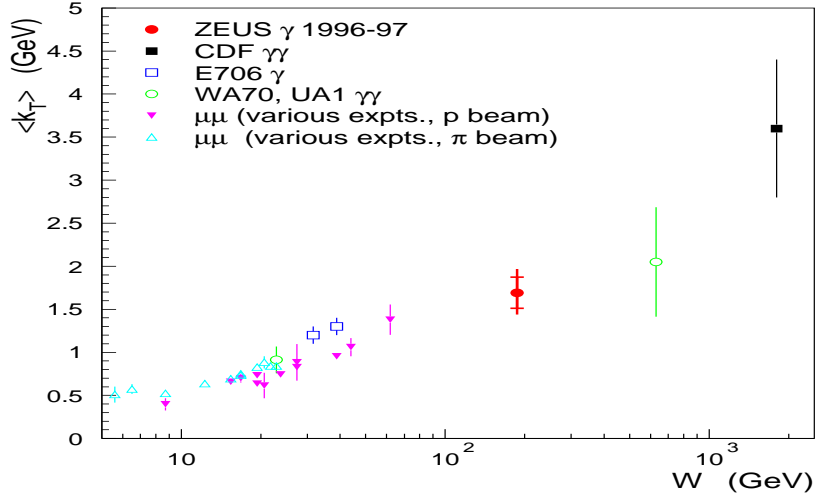,width=10cm,width=12cm,height=9cm}
   \vspace{-1.2cm}
   \caption{Measurements of $\kT$ for a number of experiments, including
            the new ZEUS result, are plotted against the centre-of-mass
            energy for the colliding particles.
            For ZEUS, these are the photon and proton.}
   \label{fig:prph_results}
\end{figure}

In Fig.\,\ref{fig:prph_results} this is plotted, together with mean $\kT$ values 
from other experiments,\cite{summary,Begel} as a function of the 
centre-of-mass energy, $W$, of the incoming particles.
It can be seen that the ZEUS measurement is consistent with the trend for
$\kT$ to rise with increasing $W$.
This rise indicates that $\kT$ should not be understood as purely intrinsic, 
due to parton confinement in the proton, but is probably the result of 
higher-order initial-state gluon radiation which increases with the phase
space available.\cite{gluons}

\section{Deeply virtual Compton scattering}

\begin{figure}[h]
   \centering
   \epsfig{file=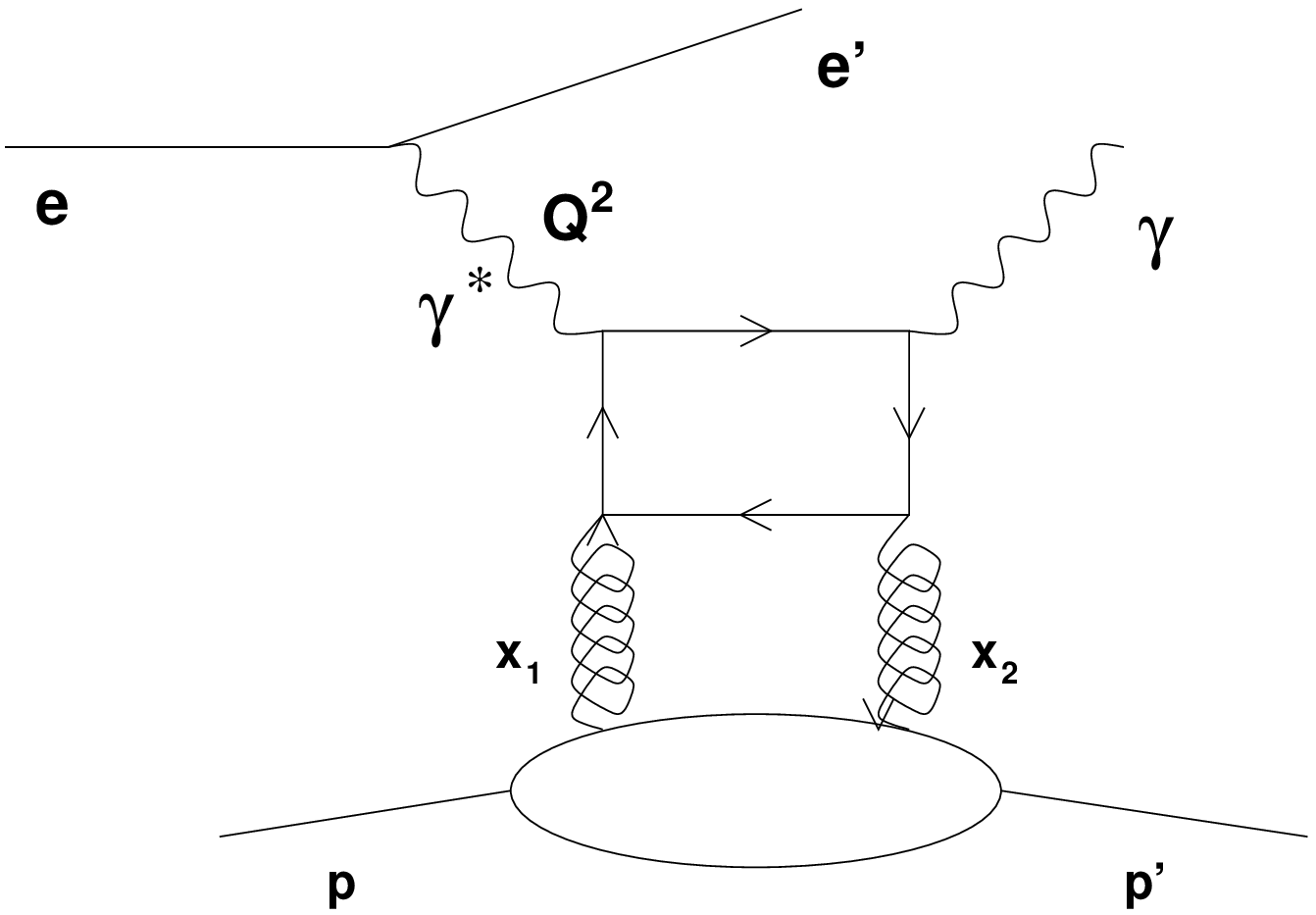,width=6cm,height=3.2cm}%
   \hspace{1.5cm}%
   \epsfig{file=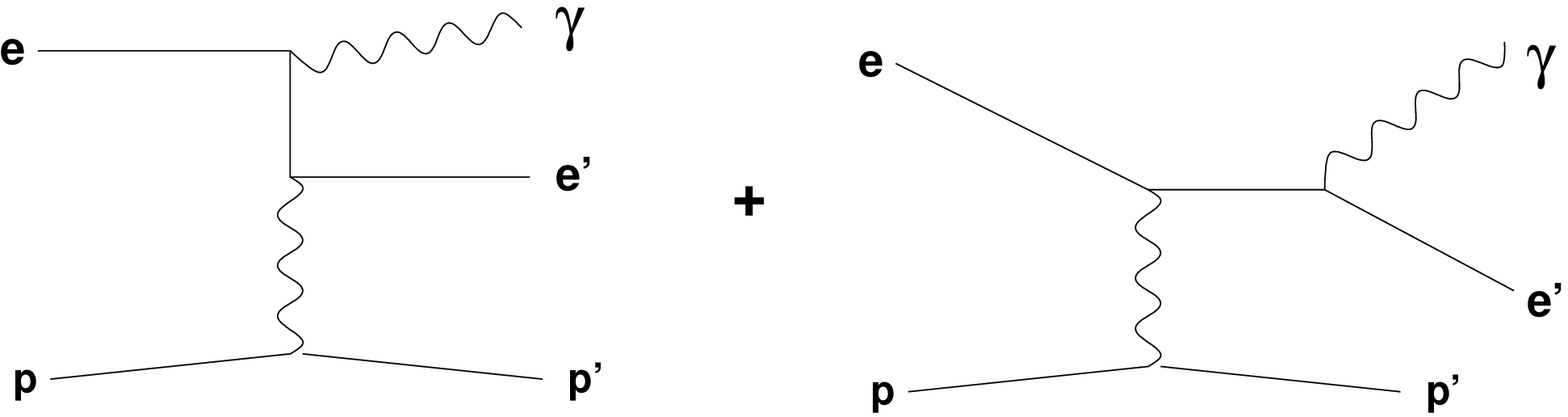,width=5.5cm,height=2.0cm,
           bbllx=0,bblly=-30,bburx=560,bbury=118}
   \caption{The quark box diagram on the left was used for the FFS 
            calculation of the DVCS cross section at low $x$.
            The right-hand diagrams are for the Bethe-Heitler process.}
   \label{fig:DVCS_diagrams}
\end{figure}

QCD-based calculations for DVCS have been performed by L.\,Frankfurt, 
A.\,Freund and M.\,Strikman (FFS)\,\cite{FFS} in terms of two gluon 
exchange with the proton as shown in Fig.\,\ref{fig:DVCS_diagrams}.
The calculations also include the purely electromagnetic Bethe-Heitler 
process, which dominates the total cross section for 
$e^+p \ra e^+\gamma p$, together with the interference term.

Both ZEUS and H1 have developed Monte Carlo event generators based on 
these calculations. 
Studies by ZEUS have shown that it is possible to separate the DVCS signal
from the Bethe-Heitler background by selecting events in which there is 
a backward positron and a photon in the central or forward region
(relative to the proton beam direction).

\begin{figure}[h]
   \centering
   \epsfig{file=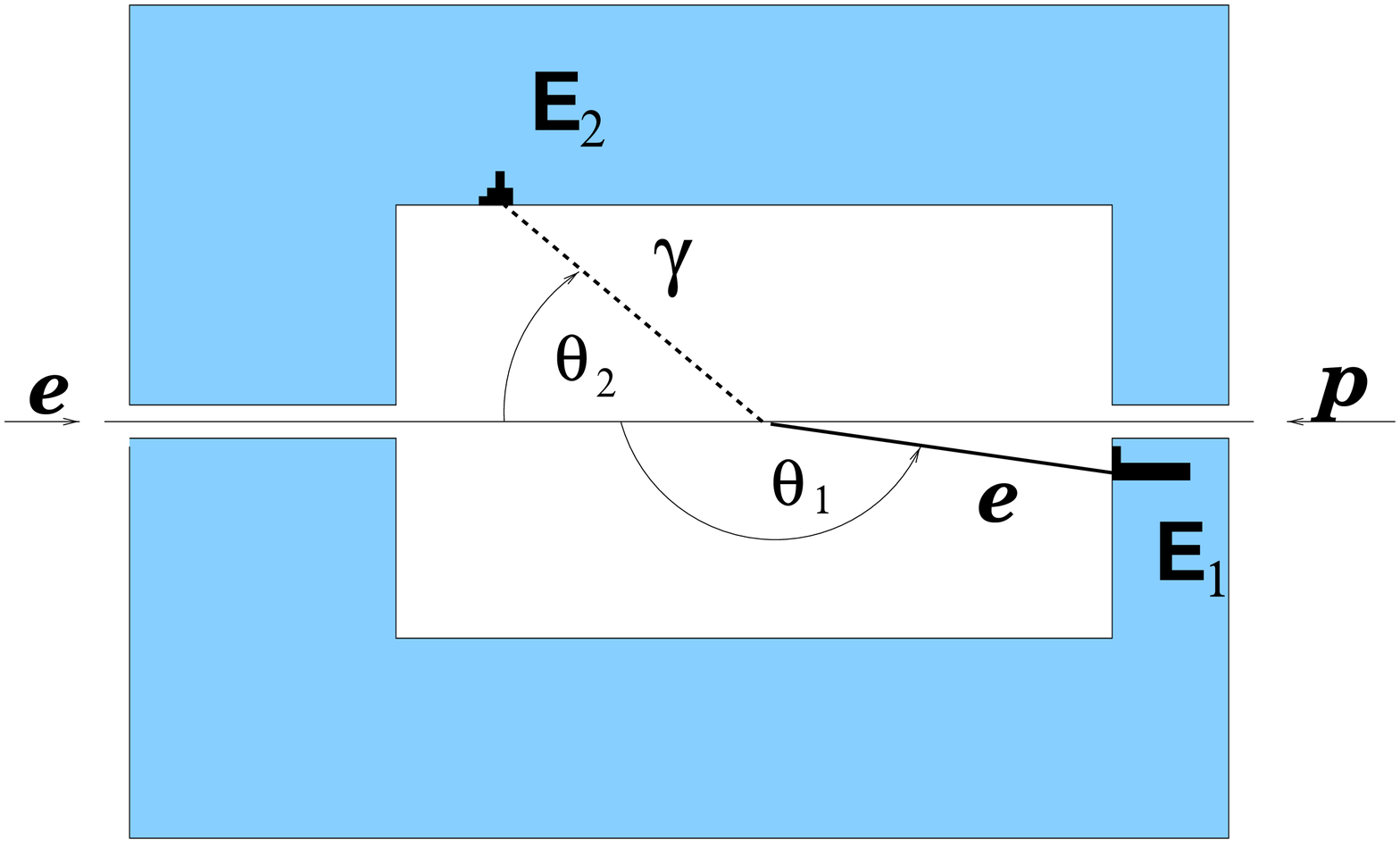,width=5.5cm,height=4.0cm,
           bbllx=0,bblly=-150,bburx=754,bbury=402}%
   \hspace{1cm}%
   \epsfig{file=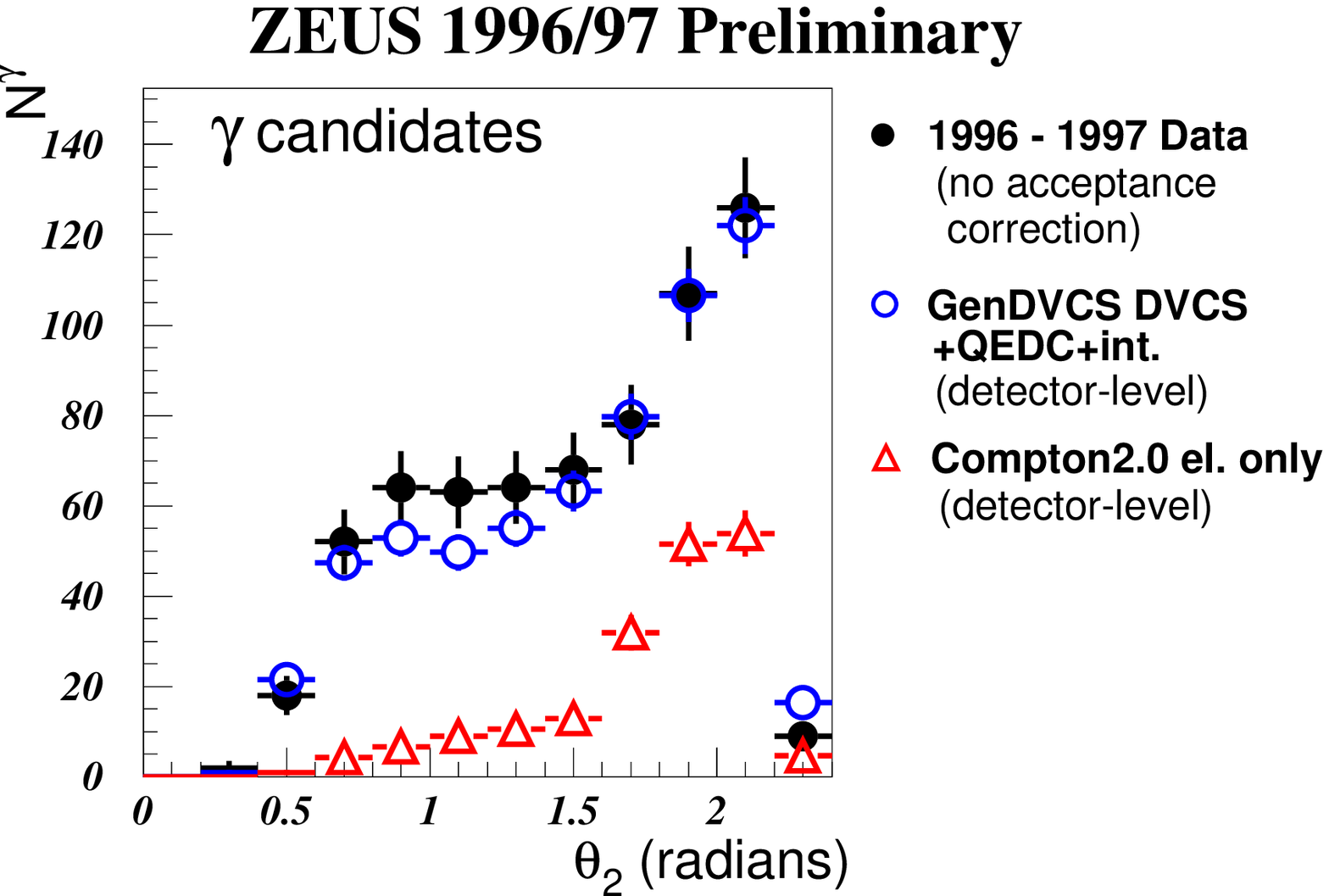,width=8.5cm}
   \vspace{-0.3cm}
   \caption{The schematic diagram on the left defines the variables used
            in the ZEUS selection of DVCS candidates.
            The right-hand plot shows the distribution of $\theta_2$
            compared to the Monte Carlo predictions for the Bethe-Heitler
            (QEDC) process with and without DVCS.}
   \label{fig:ZEUS_DVCS}
\end{figure}

The first observation of DVCS at HERA was made by ZEUS using the selection 
cuts:
$$\begin{array}{ccc}
\theta_1>2.8\rad             &  \hspace{1cm}  &  E_1>10\GeV    \\
\theta_2<2.4\rad             &  \hspace{1cm}  &  E_2>2 \GeV    \\
|\theta_1-\theta_2|>0.8\rad  &  \hspace{1cm}  &  Q^2>6\GeV^2           
\end{array}$$
where the variables are as shown in Fig.\,\ref{fig:ZEUS_DVCS}.
The photon shower shapes were studied and found to be incompatible with
$\pi^0$ and $\eta^0$ production.
As may be seen from Fig.\ref{fig:ZEUS_DVCS}, there is  a significant 
excess of events over the pure Bethe-Heitler prediction but the data 
agree well with the prediction of the DVCS plus Bethe-Heitler Monte Carlo
generator.

The analysis of data from the H1 detector, which has been described 
elsewhere,\cite{H1} was also based on the requirement for two electromagnetic 
clusters, one with an energy above $15\GeV$ in the backward calorimeter
and a second in the central part of the LAr calorimeter with $p_T>1\GeV$, 
where $p_T$ is the transverse momentum with respect to the proton beam.
To remove background, events were rejected if there were other clusters 
above $0.5\GeV$ or if signals were seen in the forward detectors.
Events were also rejected if more than one track was found, or if the
reconstructed track did not point to one of the clusters. 
If no track was seen, the backward cluster was assumed to be the positron.

For DVCS candidates, the photon was required to be in the central part
of the calorimeter and the positron in the backward part.
Events with the positron in the central part and the photon in the 
backward were used as a control sample.
This consisted almost entirely of Bethe-Heitler events, with a negligible
DVCS contribution, owing to the large scattering angle of the positron.
However, there was a small background from diffractive $\rho$ production
and from electron pairs.

\begin{figure}[h]
   \centering
   \epsfig{file=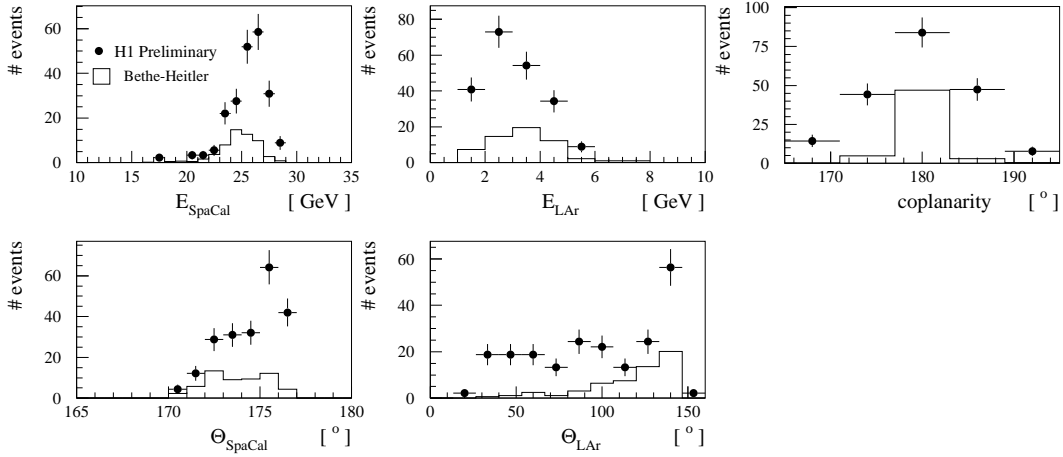,width=14cm}
   \caption{Uncorrected event distributions for the H1 DVCS candidates
            are compared to the Monte Carlo predictions for the 
            Bethe-Heitler process only.
            The distributions are of the energy and polar angle of
            the two clusters, where LAr indicates the central calorimeter
            and SpaCal the backward one, and of the event coplanarity
            (see text).}
   \label{fig:H1_plots}
\end{figure}

Figure \ref{fig:H1_plots} shows a number of distributions for the DVCS 
candidates compared to the Monte Carlo calculations for Bethe-Heitler
events only.
The large excess of events above the prediction is clear evidence for DVCS.
Note in particular that the distribution of the coplanarity (the difference
in azimuthal angle of the two clusters) is much broader than for the 
Bethe-Heitler process.
This is because the diffractive DVCS events have a much flatter $t$ 
dependence than the purely electromagnetic Bethe-Heitler process and so
have a poorer $p_T$ balance between the positron and photon.

An important check was made by studying the same distributions as plotted
in Fig.\,\ref{fig:H1_plots} for events in the control sample, instead of
DVCS candidates.
In this case there was good agreement between the data and the predictions
of the Monte Carlo generator for the Bethe-Heitler process plus background, 
so indicating that the acceptance of the detector was well understood.

\begin{figure}[h]
   \centering
   \psfig{file=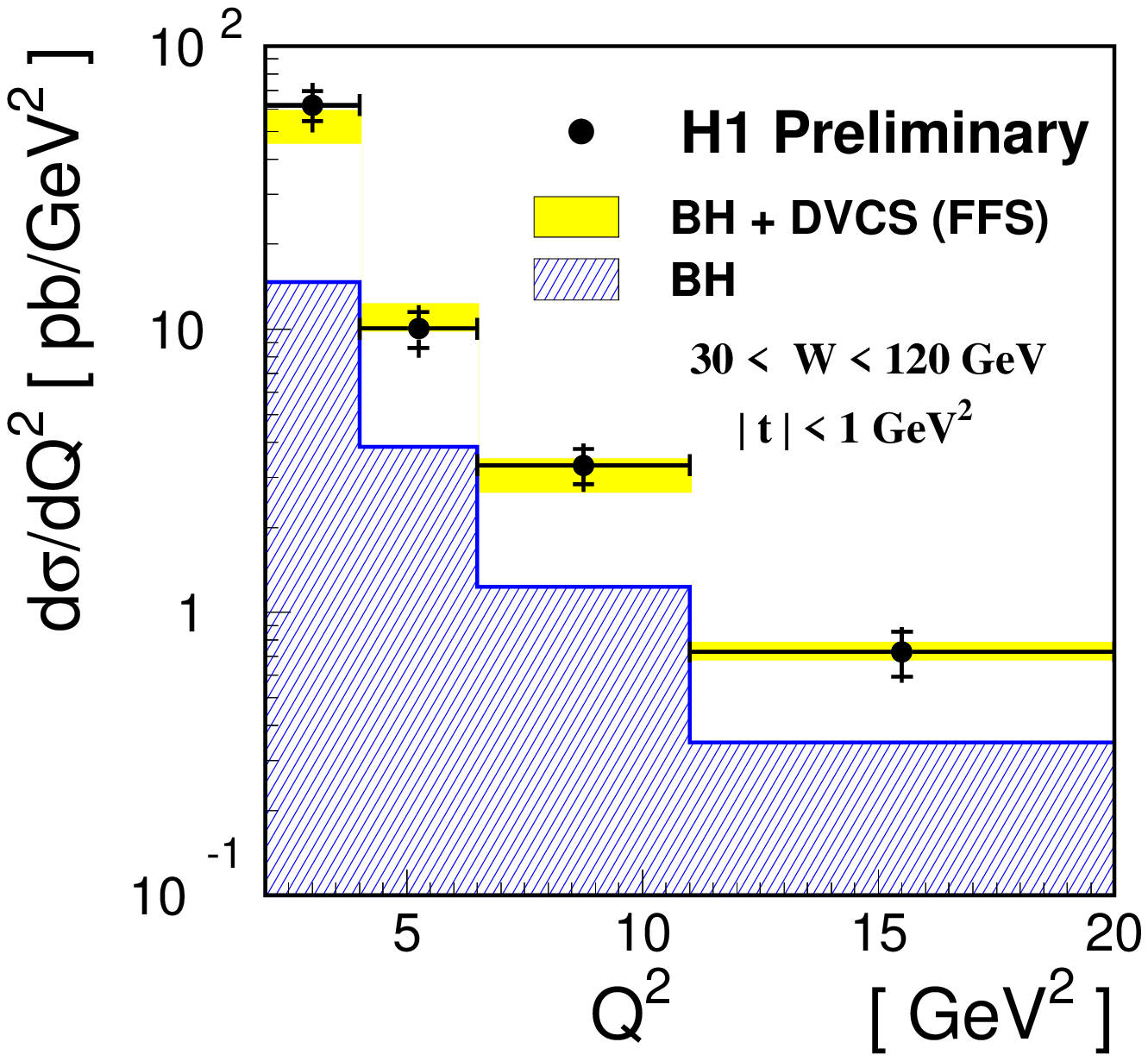,width=6.5cm,height=5.5cm}%
   \hspace{0.5cm}%
   \psfig{file=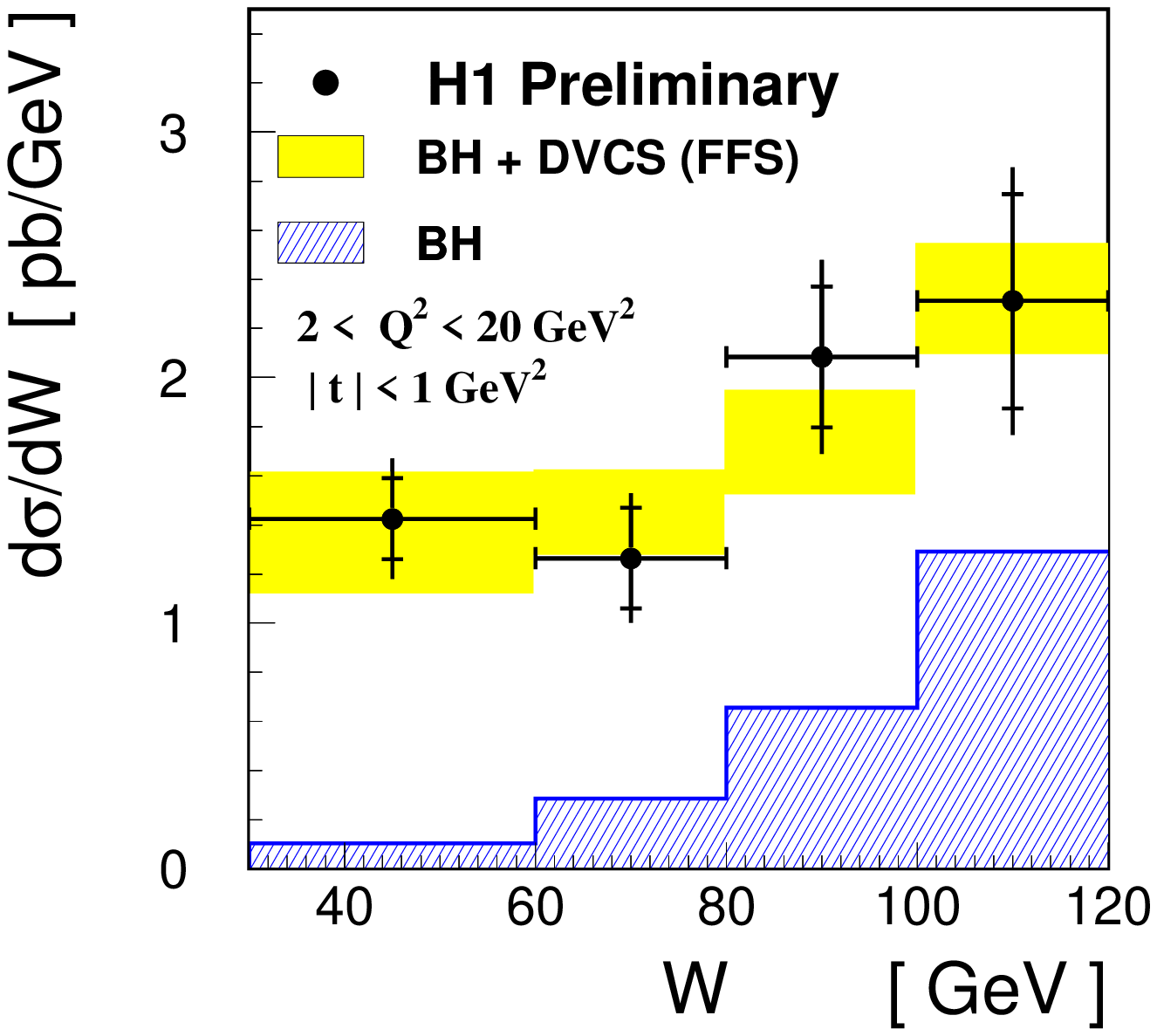,width=6.5cm,height=5.5cm}
   \vspace{-0.3cm}
   \caption{The cross sections for $e^+ p \ra e^+ \gamma p$ measured by
            H1 as functions of $Q^2$ and $W$ are compared to the FFS
            predictions.
            The predictions for the Bethe-Heitler process alone are 
            also shown.}
   \label{fig:H1_xsect}
\end{figure}

To calculate cross sections, the kinematics of the DVCS candidates was 
reconstructed by the double angle method and the Monte Carlo generator 
was used to correct for the detector acceptance.
Figure \ref{fig:H1_xsect} shows the differential cross sections as functions
of $Q^2$ and $W$.
The data are compared to the Monte Carlo predictions for the Bethe-Heitler
process and for Bethe-Heitler plus DVCS (FFS).\cite{FFS}
The range of the FFS prediction represents the uncertainty in the 
$t$ dependence of the DVCS cross section.
However, it is clear that the results are well described by the model.

\section{Conclusions}

Prompt photons provide a clean way to study QCD and photon structure.
ZEUS has used such events to make a new determination of $\kT$, the 
effective transverse parton momentum in the proton.
This is consistent with the trend seen by other experiments for $\kT$
to rise as W increases.

First signals for DVCS have been seen by both H1 and ZEUS.
H1 has also measured cross sections which are in good agreement with
QCD-based predictions.
The interference between DVCS and the Bethe-Heitler process is still to
be investigated, but the better statistics expected after the HERA upgrade
will be a great advantage.

\section*{Acknowledgments}
I should like to thank the Moriond organisers for the invitation to attend
a most stimulating conference.
I am also grateful for many helpful discussions with my ZEUS and H1 
colleagues.

\end{document}